\documentclass[aps,prb,amsmath,amssymb,nobibnotes,longbibliography,superscriptaddress,twocolumn]{revtex4-2}

\usepackage{t1enc}
\usepackage[utf8]{inputenc}
\usepackage{amsmath}

\usepackage{graphicx,bm,amsmath,amssymb,natbib,color}
\usepackage[unicode=true, colorlinks=true, citecolor={blue!80!black}, urlcolor={blue!50!black}, linkcolor = {blue!80!black}]{hyperref}
\usepackage{physics}
\usepackage{bm}
\usepackage{bbm}
\usepackage[dvipsnames]{xcolor}
\usepackage{soul}
\graphicspath{{figs/}}

\usepackage[english]{babel}

\begin{document}

\title{Plasmonic Modes in hybrid Josephson Junction Arrays}

\newcommand{\affC}{\affiliation{Centro At\'omico Bariloche and Instituto Balseiro, CNEA, CONICET, 8400 San Carlos de Bariloche, R\'io Negro, Argentina}}
\newcommand{\affR}{\affiliation{Institut für Experimentelle und Angewandte Physik, University of Regensburg, 93053 Regensburg, Germany}}
\newcommand{\affT}{\affiliation{CNR - Istituto Officina dei Materiali (IOM), 34149 Trieste (Italy)}}
\newcommand{\affCl}{\affiliation{Halle-Berlin-Regensburg Cluster of Excellence CCE, University of Regensburg, 93040 Regensburg, Germany}}

\author{Alex Kirchner} \affR\affCl
\author{Simon Feyrer} \affR\affCl
\author{Vjeko Dimi\'c} \affR
\author{Johanna Berger} \affR
\author{D. Curcio} \affT
\author{G. Biasiol} \affT
\author{Michael Prager} \affR
\author{Matthias Kronseder} \affR
\author{Dominique Bougeard} \affR
\author{Nicola Paradiso} \affR\affCl
\author{Christoph Strunk} \affR\affCl
\author{Leandro Tosi}\email[Corresponding author: ]{leandro.tosi@ib.edu.ar} \affR\affC

\date{\today}

\begin{abstract}
We present microwave characterization measurements of Josephson junction arrays (JJAs) based on epitaxial Al-InAs quantum well heterostructure. The Josephson inductance of the constituent planar junctions can be derived by probing the low-energy plasmon modes of these devices. Applying an out-of-plane magnetic field gives rise to a Fraunhofer-like diffraction pattern, from which the current-phase relation and the effective transparency can be extracted. JJAs can be used to achieve high inductances, suitable for the implementation of quantum circuits. They also provide an excellent test-bed for studying the microscopic excitations of hybrid superconductor-semiconductor devices associated with the presence of Andreev states. Here, we demonstrate their tunability over a broad range of out-of-plane magnetic fields.
\end{abstract}
\maketitle

\section{Introduction}
Superconducting quantum circuits are considered a promising route toward quantum computing \cite{krantz-2019, arute-2019, kim-2023}, while also serving as a versatile playground for the experimental study of quantum mechanics \cite{hofheinz2008generation,hofheinz2009synthesizing,ansmann-2009,oconnell-2010,vijay-2011,ofek2016extending,minev2019catch,jafferis2022traversable,google2023suppressing}. Common architectures, such as the transmon \cite{koch-2007} and the fluxonium \cite{manucharyan-2009}, rely on the tunnel Josephson junction as their distinctive nondissipative, nonlinear component \cite{devoret-1997}. However, because the Josephson effect is a general phenomenon ubiquitous in superconducting weak links \cite{kulik-1969,likharev-1979,Blonder1982,datta-1996}, there is growing interest in integrating mesoscopic Josephson junctions into quantum circuits \cite{pita-vidal-2025pre}. While initial efforts focused on testing the mesoscopic Josephson effect across a great variety of weak links, ranging from atomic contacts \cite{dellarocca-2007,bretheau-2013-exciting}, carbon nanotubes \cite{Kasumov1999,JarilloHerrero2006,pillet-2010}, and graphene flakes \cite{Bretheau2017}, to semiconducting nanowires \cite{Doh2005,goffman-2017,vanWoerkom2016,Spanton2017}, quantum dots \cite{Deacon2010,Lee2014}, topological insulators \cite{Murani2017,Bocquillon2016,Wiedenmann2016,Ren2019} and two-dimensional electron gases \cite{shabani-2016, kjaergaard-2015thesis,kjaergaard-2017,elfeky-2024}, the past decade has been devoted to probing their internal degrees of freedom. Such endeavors require few-channel, high-transparency tunable junctions. The coherent manipulation of Andreev bound states (ABS) \cite{janvier-2015, Hays2018}, realized using cQED techniques \cite{blais-2021}, demonstrated a new type of mesoscopic qubit: the Andreev qubit \cite{desposito-2001, chtchelkatchev-2003}, and established the grounds for the quantum control of the internal energy levels of Josephson junctions. Intrinsic properties introduced by different weak links, such as spin-orbit coupling, contribute to the richness of the mesoscopic Josephson effect. In InAs nanowires, for instance, ABS are imprinted with a spin texture that can be revealed by microwave spectroscopy \cite{tosi-2019,bargerbos-2020} and used as a platform for a superconducting spin qubit \cite{padurariu-2010,hays-2021,pita-vidal-2023}.

Currently, there are four primary directions driving the development of hybrid quantum circuits: (i) the engineering of novel circuit components, (ii) the design and exploration of distinct potential landscapes, (iii) the realization of Andreev-flavored bosonic qubits, and (iv) the development of fermionic-bosonic qubits. A highly promising avenue involves mesoscopic junctions to build new components or imprint novel properties into standard architectures. For instance, embedding gate-tunable junctions \cite{phan-2023, hao-2024, strickland-2023,splitthoff-2024, butseraen-2022} or SQUIDs \cite{scherubl-2025} into microwave resonators provides \textit{in situ} tunability, enabling new designs for parametric amplifiers and dynamically tunable nonlinearities. Mesoscopic junctions are also inherently suitable for the realization of the superconducting diode effect \cite{baumgartner-2021, baumgartner-2022}. Expanding beyond two terminals, multi-terminal Josephson junctions offer pathways for non-reciprocal devices such as on-chip quantum circulators \cite{virtanen-2024}. Furthermore, the evolution of semiconductor-based qubits, from the gate-tunable transmon or ``gatemon'' (realized using nanowires \cite{larsen-2020,deLange2015,casaliglesias2026ultrastrong}, 2DEGs \cite{casparis-2018}, 2DHGs \cite{Sagi2024}, carbon nanotubes \cite{riechert-2025}, and graphene \cite{Wang2019} as weak links) to the gate-tunable fluxonium or ``gatemonium'' \cite{pita-vidal-2020,strickland-2025}, highlights the immense potential of mesoscopic junctions. Simultaneously, hybrid weak links allow for the engineering of unique energy-phase relations. This includes the realization of $\cos(2\varphi)$ elements \cite{gyenis-2021,feldstein-bofill-2026pre} for hardware-protected qubits, ballistic ABS \cite{vakhtel-2023, vakhtel-2024}, and the exploitation of distinct parity states, such as the odd-even qubit \cite{bargerbos2022, Sahu2024}. Beyond pure energy landscapes, the internal degrees of freedom in these junctions are blurring the lines between fundamental qubit modalities. This is evidenced by Andreev-flavored bosonic qubits, pioneered by proposals from Saclay \cite{caceres2026ferbo} and recent multi-terminal implementations \cite{matute-canadas-2024}, which offer protection at the hardware level. Another example are fermionic-bosonic qubits, where the spin degree of freedom of ABS is intertwined with the wave function of the bosonic modes of the circuit \cite{pita-vidal-2023,Bargerbos2023}. Realizing this vast potential, particularly the leap to complex arrays required for gatemonium \cite{strickland-2025} and protected regimes, demands a robust, highly transparent, and tunable platform, like that provided by two-dimensional Al/InAs heterostructures.

To characterize these mesoscopic junctions and realize advanced circuit proposals, accurate microwave spectroscopy is essential. Microwave resonators provide a reliable means to extract the Josephson inductance, $L_J$. Cold off-chip resonators coupled to the sample offer a robust approach, as the bare resonator remains largely insensitive to environmental parameters such as temperature and magnetic field \cite{fuchs-2022, baumgartner-2021, baumgartner-2022}. However, observing sizable dispersive shifts in this geometry requires a massive junction inductance, as resolution is typically constrained by resistive losses and the lower operating frequencies dictated by macroscopic components. Conversely, on-chip superconducting resonators benefit from lower losses and higher operating frequencies \cite{watanabe-1994, strickland-2023, tanaka-2025, ruggiero-2026pre}. Yet, they face significant drawbacks when implemented with the thin superconducting films ubiquitous in superconductor-semiconductor heterostructures. First, the bare resonance frequency becomes strongly dependent on magnetic field and temperature due to variations in the kinetic inductance of the film \cite{bothner-2017, phan-2023, hao-2024, strickland-2023, scherubl-2025, medahinne-2025, ruggiero-2026}. Extracting the underlying $L_J$ requires carefully decoupling these background effects. Second, standard planar resonators degrade rapidly in out-of-plane magnetic fields; the quality factor is drastically suppressed primarily by dissipation from moving Abrikosov vortices, alongside quasiparticle excitations \cite{fuchs-2022}. 

In this work, we address these limitations by utilizing a resonator design in which a Josephson junction array (JJA) serves as the center conductor of a coplanar waveguide (CPW) \cite{castellanos-beltran-2007, weissl-2015, kuzmin-2019, mukhopadhyay-2023, bubis-2026, oh-2026, manucharyan-2009}. By probing the low-energy plasmonic modes of the device, we extract the out-of-plane magnetic-field dependence of the Josephson inductance. This architecture offers three distinct advantages. First, the total Josephson inductance is exceedingly large, completely dominating over the kinetic inductance of the thin film. The resonance frequency is thus directly governed by the junctions, scaling as $f \propto 1/\sqrt{L_J}$. Second, because the array segments the superconducting film into sub-micron islands, the nucleation and flow of vortices in the center conductor is severely restricted. Finally, the massive inductance increases the characteristic impedance of the circuit; consequently, residual losses have a reduced impact on the internal quality factor. Notably, this robust high-inductance architecture allows us to resolve the junctions' interference pattern in a broad range of magnetic fields, revealing multiple higher-order sidelobes. Ultimately, this component has the potential to operate as a superinductor for the realization of protected quantum circuits \cite{gyenis-2021,vakhtel-2024,caceres2026ferbo, matute-canadas-2024}.

\section{Planar Josephson junction arrays in Al/InAs}
\begin{figure}[t!]
    \centering
\includegraphics[width=1\columnwidth]{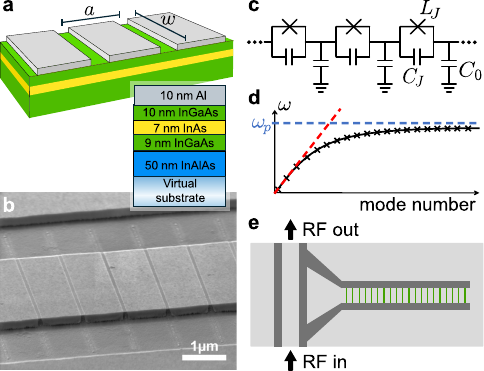}
    \caption{\label{fig1} 
    \textbf{Device layout.} \textbf{a}, Schematic of the Josephson junction array (JJA) based on a proximitized 2DEG. The inset shows the heterostructure of device A. The superconducting top layer is depicted in gray, the 2DEG in yellow. \textbf b, Scanning electron micrograph of one of the devices, taken prior to the deposition of the AlOx dielectric. \textbf c, Circuit model used to describe the electrodynamics of the array. The Josephson junctions are modeled by their Josephson inductance $L_J$. \textbf d, Dispersion relation of the circuit model assuming $C_0/C_J=2/3$. The red line indicates the linear regime for small mode number, the blue line indicates the plasma frequency. The black crosses show the expected resonances for a JJA consisting of 20 junctions. \textbf e, Schematic of the JJA in the $\lambda/4$-resonator configuration. The array is connected to ground on the right side and capacitively coupled to a transmission line on the left side. }
\end{figure}
Our devices are fabricated using an Al/InAs heterostructure, comprising a near-surface InAs-based 2DEG, which is proximitized by a thin layer of epitaxially grown aluminum \cite{shabani-2016, zhang-2023,Kirti2025} (see Fig.~\ref{fig1}a). We present results of two devices: A and B, fabricated from similar heterostructures. Junctions are realized by selectively etching a gap with a length of $\sim$100\,nm in the top aluminum layer. The JJAs consist of a chain of Josephson junctions with spacing $a=1\,\mathrm{\mu m}$, as shown in Fig.~\ref{fig1}a,b. The width of the junctions is $w=3.1\,\mathrm{\mu m}$. 
The array serves as the center conductor of a coplanar waveguide, forming a $\lambda/4$-resonator. The circuit model of the device is shown in Fig.~\ref{fig1}c, consisting of a capacitance between adjacent superconducting islands $C_J$ and a capacitance $C_0$ to the ground plane. The dispersion relation $\omega(k)$ of plasmon modes in JJAs has been thoroughly discussed in Ref.~\cite{weissl-2015}. In the long wavelength, low frequency limit, the dispersion relation is linear (red line in panel d). At frequencies approaching the plasma frequency of a single junction $\omega_p=1/\sqrt{L_JC_J}$ (blue line in d), the dispersion deviates from the linear regime. Compared to many implementations of JJA resonators \cite{ bothner-2017,phan-2023, hao-2024, strickland-2023,scherubl-2025,medahinne-2025,ruggiero-2025pre}, in our device the distance between the JJA and the surrounding ground plane is small ($\approx 2\,\mathrm{\mu m}$). Thus, $C_0$ is larger and becomes comparable to $C_J$. Furthermore, $C_J$ is small due to the planar geometry, resulting in a high plasma frequency on the order of $1\,\mathrm{THz}$, well above the accessible frequency range. Consequently, all measurements are performed in the linear regime of $\omega(k)$.

The JJA is capacitively coupled to a readout transmission line (see Fig.~\ref{fig1}e). The coupling is intentionally chosen to be relatively large, ensuring that the resonance remains observable even at high magnetic fields, where the internal quality factor is strongly suppressed. The devices are characterized by microwave transmission spectroscopy using a vector network analyzer. At the resonance frequencies of the JJA, a dip in the transmission coefficient $S_\text{21}$ is observed. 

\begin{figure}
    \centering
    \includegraphics[width=\columnwidth]{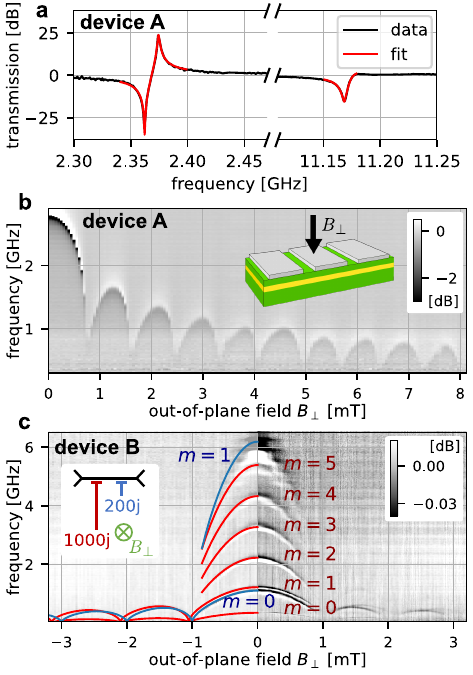}
        \caption{\textbf{Transmission measurement and interference pattern.} \textbf a, Transmission trace of device A at zero magnetic field. The two resonances are fitted by the formula given in \cite{deng-2013}. The background is subtracted using a reference measurement at higher temperature and magnetic field. \textbf b, Interference pattern of the same device. The colorbar shows the transmission parameter $S_{21}$ with the background subtracted as in panel a. \textbf c, Interference pattern of device B. In this device, two JJA resonators with $1000$ and $200$ junctions, respectively, are coupled to the same transmission line. Thus modes of both devices are visible in the colorplot. The red and blue line in the left half are guides to the eye and indicate the modes corresponding to the $1000$ and $200$ junction devices, respectively. The color scale represents the difference in transmission $S_\mathrm{21}$ measured at two different input powers. This differential measurement was employed to enhance the visibility of the resonances. }
    \label{fig2}
\end{figure}

\section{Probing the Josephson inductance}
The devices are measured in a dilution refrigerator at temperatures spanning $T\approx 20-100\,\mathrm{mK}$. Fig.~\ref{fig2}a shows the transmission spectrum of the array. The first two harmonics are visible in the accessible frequency range of our setup. The resonances are strongly asymmetric and can be fitted with the formula given in Ref.~\cite{deng-2013}.
The spacing of the modes deviates from the expected spacing for a $\lambda/4$-resonator. This can be attributed to the strong coupling to the read-out transmission line, which modifies the boundary condition. The shift of the resonance frequencies can be reproduced using a model including the coupling capacitance, discussed in Appendix \ref{ap:strongCoupling}. Fitting the model parameters to the measured resonance frequencies yields a value for both the coupling capacitance $C_C$ and the bare resonance frequency
\begin{equation}
    \omega_0 = \frac{1}{N\sqrt{ L_JC_0}}, 
\end{equation}
where $N$ is the number of junctions. 
The capacitance of one island to ground $C_0$ can be calculated using conformal mapping techniques \cite{watanabe-1994}. At zero field, Josephson inductances of $180\,\mathrm{pH}$ and $1.5\,\mathrm{nH}$ per junction were determined for devices A and B, respectively. However, this value is subject to a large systematic error ($\sim 20\%$), as it depends strongly on the exact geometry \footnote{Additionally, in device A, 18 of the 300 junctions were not fabricated correctly due to lithography issues and are missing. Since the corresponding islands still contribute to the capacitance, the formula $\omega_0=1/\sqrt{N_JL_J N_\text{islands}C_0}$ was used to calculate the Josephson inductance, with $N_J=282$ and $N_\text{islands}=300$.}. 

While it is difficult to reliably extract an absolute value of the Josephson inductance, the relative shift in magnetic field remains unaffected by this uncertainty. Assuming that other contributions to the inductance are negligible, the resonance frequency is proportional to the inverse square root of the Josephson inductance $f\propto 1/\sqrt{L_J}$. This is a good approximation because both the geometric and kinetic inductance contributions are expected to be at least a factor $100$ smaller than the Josephson inductance. As discussed in Appendix \ref{ap:strongCoupling}, this proportionality is not affected by strong coupling effects.

\section{Anomalous Fraunhofer pattern}

\begin{figure*}[t!]
    \centering
    \includegraphics[width=1\textwidth]{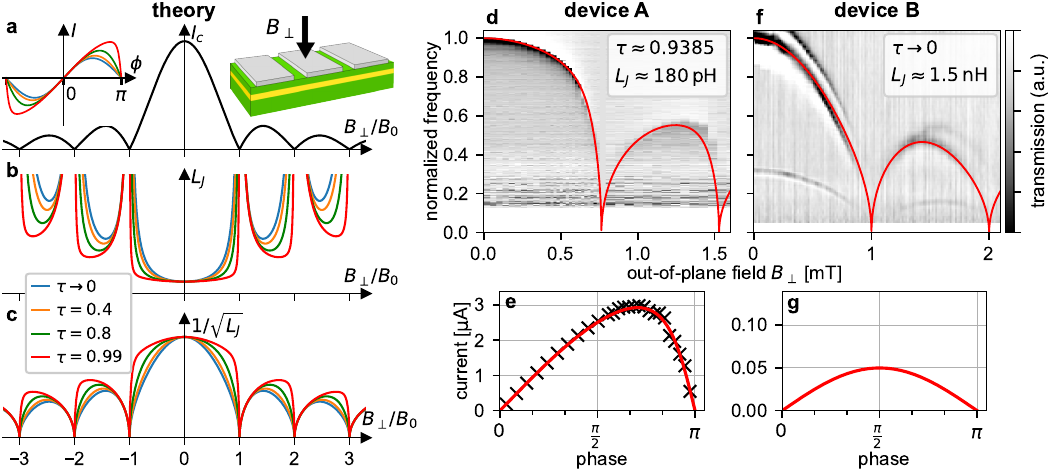}
    \caption{
    \textbf{Extracting the CPR from the measurement of the interference pattern.}
    \textbf{a,} 
    Interference pattern of the critical current, assuming a Beenakker-Furusaki type current phase relation, as shown in the inset. It is numerically identical for all values of the transparency $\tau$. 
    \textbf{b,} Interference pattern of the Josephson inductance for different values of the transparency $\tau$.  
    \textbf{c,} Interference pattern of the inverse squared Josephson inductance. 
    \textbf{d,} Colorplot of the measured interference pattern of device A. A fit of the first lobe is shown in red. 
    \textbf{e,} Extracted CPR of device A. The resonance curves within the first lobe are fitted (black crosses) and fitted to the Beenakker-Furusaki form.
    \textbf{f,} Colorplot of the measured interference pattern of device B. Here the resonances could not be fitted due to their poor quality. A sinusoidal CPR is assumed, and its periodicity is chosen to match that of the interference pattern. Below $\tau=0.3$, there is no significant deviation from the sinc-shape.
    \textbf{g,} Expected sinusoidal CPR for device B.
     }
    \label{fig3}
\end{figure*}

The following measurements are performed in out-of-plane magnetic field. This introduces hysteresis and abrupt jumps of the resonance frequency due to vortex trapping. To mitigate these effects, the sample is heated above the critical temperature of the aluminum before each measurement at a given magnetic field value. 

Applying an out-of-plane magnetic field leads to interference effects, producing a Fraunhofer-like interference pattern. The pattern manifests as a modulation of the resonance frequency as shown in Fig.~\ref{fig2}b. At integer flux quanta threading the effective junction area, the critical current of each junction is reduced and the Josephson inductance increases. This can be observed as a decrease in the resonance frequency. Owing to the high participation ratio $L_J/(L_\text{kin}+L_\text{geo})$, this modulation is substantial and reaches almost zero. Fig.~\ref{fig2}c shows the interference pattern of device B, comprising two JJAs connected to a single transmission line. Resonances of both JJAs are visible in the spectrum. In addition, the higher Josephson inductance of these junctions enables the direct observation of multiple harmonics of the JJAs.

To model the interference effect we start by recalling that the Josephson inductance is determined by the derivative of the current-phase relation (CPR) at its equilibrium phase and is given by 
\begin{equation}
    L^{-1}_J = \frac{2\pi}{\Phi_0} \frac{\mathrm d I(\phi)}{\mathrm d\phi}.
\end{equation}
To calculate $I(\phi)$ with out-of-plane magnetic field, we assume that the phase difference varies linearly along the width of the junction $\phi(x,\gamma) = \gamma + \frac{2\pi B_\perp}{B_0} \frac{x}{w}$. Here, $B_0$ denotes the magnetic field corresponding to one flux quantum threading the effective area of the junction, and $\gamma$ is the gauge-invariant phase difference. The different paths of the current across the junction experience different phase differences and therefore interfere. The total supercurrent results from the integration of the local current density $j(\phi(x,\gamma))$ along the width of the junction: 
\begin{equation}\label{eq:integration_total_current}
    I_\Sigma(\gamma)=\int_{-\tfrac{w}{2}}^{\tfrac{w}{2}}\mathrm dx\, j(\phi(x,\gamma)).
\end{equation} 
Then the critical current can be obtained as $I_c(B_\perp)=\max_{\gamma}(I_\Sigma(\gamma))$, and the Josephson inductance as $L^{-1}_J(B_\perp)=(2\pi/\Phi_0) ({\mathrm d I_\Sigma(\gamma)}/{\mathrm d\gamma})$. 

For a sinusoidal CPR, this gives
\begin{equation} 
    I_\Sigma(\gamma)=I_0\cdot\left[\frac{\sin\left(\pi\frac{B_\perp}{B_0}\right)}{\pi\frac{B_\perp}{B_0}}\right]\sin(\gamma),
\end{equation}
thus the critical current, $I_c(B_\perp)$ exhibits the well-known sinc-shaped Fraunhofer interference pattern (see Fig.~\ref{fig3}a). In this case, the Josephson inductance is simply inversely proportional to the critical current. For a non-sinusoidal CPR, the shape of the interference pattern can differ from this sinc-shape. Baumgartner \textit{et al.} \cite{baumgartner-2021} demonstrated that for high transparency SNS junctions, where the CPR has higher harmonics, the interference pattern in Josephson inductance is sensitive to the shape of the CPR.

In SNS junctions, the Josephson current is carried by Andreev bound states, and the CPR results from the contribution of the different conduction channels. For a single channel of transparency $\tau$ at $T=0$ it is given by the Beenakker-Furusaki formula: \cite{beenakker-1991, dellarocca-2007, golubov-2004, baumgartner-2021}
\begin{equation}\label{eq:beenakker}
	I(\phi)=wj(\phi) = \frac{e\Delta^{\!*}}{\hbar}\frac{{\tau}\sin(\phi)}{2\sqrt{1-{\tau}\sin^2(\tfrac{\phi}{2})}},
\end{equation}
where $\Delta^*$ is the superconducting gap induced in the InAs quantum well underneath the aluminum film \cite{baumgartner-2021}. The inset in Fig.~\ref{fig3}a shows the CPR from Beenakker-Furusaki for different values of $\tau$. As we discuss in the Appendix, numerical integration of $j(x,\gamma)$ shows that the interference pattern in critical current does not deviate from the sinc-shape irrespective of the transparency $\tau$. On the contrary, the interference pattern in Josephson inductance evolves from an inverted sinc-shape to a more rectangular form (see Fig.~\ref{fig3}b). At higher transparency, the interference pattern gets skewed: the central lobe becomes less rounded, and the side lobes develop an asymmetry. As seen in Fig.~\ref{fig3}c this anomalous Fraunhofer pattern in Josephson inductance translates to the modulation of the resonance frequency, and in the case of our experiments, to the modulation of the low energy modes of the JJA. In Fig.~\ref{fig2}b and c we show how the two devices A and B, respectively, illustrate this nicely. 

\section{Extracting the CPR}
In the following we show how, in the case of high-transparency SNS junctions, the fit of the Fraunhofer pattern in frequency, translated into Josephson inductance, allows us to determine the CPR.

Assuming that the CPR is antisymmetric about $\phi = 0$: $I(\phi) = -I(-\phi)$, the pattern within the central lobe is given by (see details in Appendix~\ref{ap:interference})
\begin{equation}\label{eq:LjInterferencePattern}
    L_J(B_\perp) = \frac{\hbar}{2e}\frac{\pi \frac{B_\perp}{B_0}}{I\left(\pi\frac{B_\perp}{B_0}\right)}.
\end{equation}

Hence, the CPR can be extracted directly from the shape of the central lobe, as demonstrated in Fig.~\ref{fig3}d-g. Panels d, f show the measured interference patterns of devices A and B, respectively. Panels e, g show the corresponding extracted CPR. In device A, the CPR is well described by the Beenakker-Furusaki form, Eq.~(\ref{eq:beenakker}), characterized by a single transparency parameter $\overline\tau$. Since the CPR arises from contributions of many channels with different transparencies, we interpret this parameter as an effective transparency. The good agreement suggests that most of the individual channel transparencies are narrowly distributed around their average value. This observation is consistent with the findings of Ref.~\cite{baumgartner-2021} in a similar device. For the measurement shown in Fig.~\ref{fig3}d,e, a curve fit results in a value of the effective transparency of $\overline{\tau}\approx 0.93\pm 0.06$. While the fit procedure in principle allows for an accuracy better than $0.1\%$, here the error results from the accuracy of the offset field. A more detailed discussion is provided in Appendix \ref{ap:fittingTau}.

In contrast, device B exhibits a nearly sinusoidal CPR, consistent with low transparency \footnote{The heterostructure is similar to that of device A, shown in the inset of Fig. \ref{fig1}a. The main difference is in the growth of the aluminum film, which was not performed with active cooling of the substrate. We assume that the low transparency originates from the quality of the aluminum film and its interface. This also likely explain the poor quality of the resonances.}. Due to the poor quality of the resonances, it was not possible to fit the shape of the interference pattern. The shown curve in Fig.~\ref{fig3}f is assuming a sinusoidal CPR. No significant deviations from this shape can be observed.

As the total Josephson inductance is given by the expression $L_J = (\Phi_0/2\pi)^2\cdot 4/(\tau\Delta^{\!*}N)$, the quantity $\Delta^{\!*}N\approx 16\,\mathrm{meV}$ can be extracted from the measured zero-field Josephson inductance and the determined effective transparency. Based on measurements on similar heterostructures, we assume an induced superconducting gap of $\Delta^{\!*}\approx 130\,\mathrm{\mu eV}$ \cite{baumgartner-2021, reinhardt-2024, scherubl-2025}. Using this value, we can estimate the number of channels $N\approx 123$. 


\section{Discussion and Conclusions}
In conclusion, we have demonstrated that microwave spectroscopy of Josephson junction arrays (JJAs) provides a powerful and robust tool for characterizing the mesoscopic properties of superconductor-semiconductor heterostructures. By embedding an Al/InAs JJA as the center conductor of a coplanar waveguide resonator, we successfully mitigated the severe dissipation typically caused by Abrikosov vortex motion in thin films subjected to out-of-plane magnetic fields. This architecture allowed us to resolve multiple higher-order sidelobes of the quantum interference pattern in a broad range of magnetic fields.

Crucially, because the resonance frequency is completely dominated by the macroscopic Josephson inductance of the array, we were able to directly map the supercurrent interference pattern to the underlying CPR of the junctions. For our high-transparency devices, the extracted CPR deviates significantly from the standard sinusoidal behavior and is remarkably well described by the Beenakker-Furusaki formalism. From this geometry, we extracted an effective conduction channel transparency of $\overline{\tau} \approx 0.93$, indicating a highly homogeneous, nearly ballistic transport regime across approximately $123$ parallel conduction channels. 

Beyond serving as a precise diagnostic tool for mesoscopic weak links, these results highlight the immense practical potential of 2DEG-based JJAs. The combination of high kinetic inductance, gate-tunability, and resilience to moderate magnetic fields makes this material platform an ideal candidate for realizing tunable superinductors. In our device, the total inductance was 51\~ nH and the characteristic impedance was 1.1\~k$\Omega$ by design. It can be further increased by increasing the distance to the ground plane. Such components are essential for the next generation of hardware-protected quantum circuits, including the gatemonium, and for exploring exotic topological superconducting phases where out-of-plane magnetic fields are a strict prerequisite.

\appendix
\renewcommand{\thefigure}{\Alph{section}.\arabic{figure}}

\section{Extracting the Josephson inductance in the presence of strong coupling}\label{ap:strongCoupling}
\setcounter{figure}{0}

Owing to the large coupling, the boundary condition of the JJA resonator at the capacitive coupling cannot be assumed perfectly open. Thus the resonance frequencies deviate significantly from those expected from a perfect $\lambda/4$ resonator. In order to calculate the Josephson inductance from the measured resonance frequency, this shift has to be taken into account. For this purpose, we assume a circuit model, where the JJA resonator is described by a lossless waveguide with propagation constant $\beta = \omega N\sqrt{L_JC_J}$ \cite{pozar-2012book}. Its input impedance is given by
\begin{equation}
    Z_\text{in} = iZ_r\tan(\beta l) = iZ_r\tan\left(\frac{\omega}{\omega_0}\right)
\end{equation}
The waveguide is coupled to the transmission line with a coupling capacitor $C_C$ with impedance $Z_C = (i\omega C_C)^{-1}$. 
In this circuit model, the condition for resonance is that the real part of the total impedance of the wavequide and the capacitor is zero $Re(Z_\text{in}+Z_C)=0$. For fixed $\omega_0$ and $C_C$, the resoanance frequencies can be computed numerically. 
The strong coupling leads to a shift of all modes to lower frequencies. The shift of each mode is different. Thus their spacing differs from that expected from a $\lambda/4$-resonator, as seen in Fig. \ref{fig2}a. This property can be used to experimentally determine the value of the coupling capacitor. For this purpose, the resonance frequencies of the first two harmonics are measured. The parameters $\omega_0$ and $\omega_C = Z_rC_C$ are then fitted such that the resonance frequencies calculated from these parameters match the experimentally determined values.

Evaluating the condition of resonance $Re(Z_\text{in}+Z_C)=0$ further, we find:
\begin{equation}
    \tan\left(\omega \sqrt{L_JC_J}N\right) = \frac{\sqrt{C_J}}{\omega\sqrt{L_J} C_C}.
\end{equation}
Thus, the condition is still fulfilled for the same value of $\omega\sqrt{L_J}$. This means that the relation $\omega\propto 1/\sqrt{L_J}$ is still valid in the case of strong coupling. This is a very useful property for the evaluation of the out-of-plane field dependence of the devices.


\section{Interference pattern for non-sinusoidal CPR}\label{ap:interference}
\setcounter{figure}{0}

\begin{figure}[h!]
	\centering
	\def\svgwidth{0.7\columnwidth}
	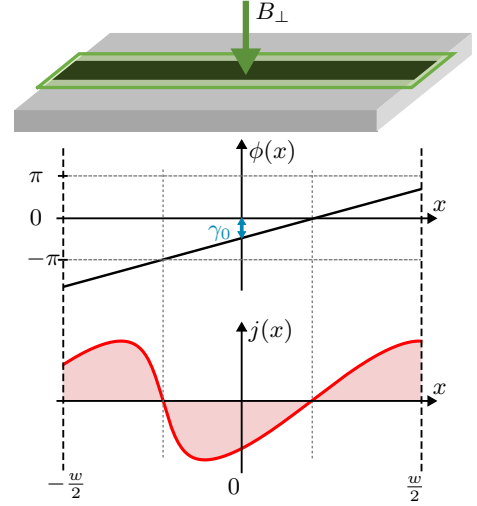
	\caption{Schematic of the interference effects of a wide junction with width $w$. An out-of-plane magnetic field $B_\perp$ is penetrating the area marked by the green rectangle. This leads to a gradient of the phase difference $\phi$ and correspondingly variations of the current density $j$ along the width of the junction.  }
	\label{fig:fraunhofer1}
\end{figure}

The magnetic field leads to variations of the phase difference along the width of the junction, as shown in Fig. \ref{fig:fraunhofer1}. The phase difference at a specific position can be obtained by integrating the vector potential along a path around the junction
\begin{equation}
	\phi(x, \gamma) = \gamma+\int\!\mathrm d\mathbf{l}\, \mathbf A = \gamma + \frac{2\pi B_\perp}{B_0}\frac{x}{w},
\end{equation}
where $\gamma$ is the gauge invariant phase difference across the junction and
$B_0 = \Phi_0/(l_\text{eff}w)$ is the magnetic field at which one flux quantum is penetrating through the effective area of the junction. As discussed by C. Baumgartner et al. \cite{baumgartner-2022}, the effective length of the junction is given by the array lattice constant $a$. In general, the effective length can be expressed as $l_\text{eff}=l+2\lambda_\text P$, with $\lambda_\text P$ the Pearl penetration length of the aluminum film which is estimated to be of the order of $20\,\mathrm{\mu m}$ in a similar heterostructure \cite{fuchs-2021thesis}. As $a<2\lambda_\text P$, the effective length is limited by the lattice spacing. Furthermore, the effective length can be also limited by the length scale associated to Andreev reflection, which is set by the coherence length $\xi$ estimated to be approximately $1\,\mathrm{\mu m}$, as discussed in \cite{Berger2026}. Since $a<2\xi$, the lattice spacing remains the relevant length scale.

\begin{figure}[h!]
	\centering
	\includegraphics[width=1\columnwidth]{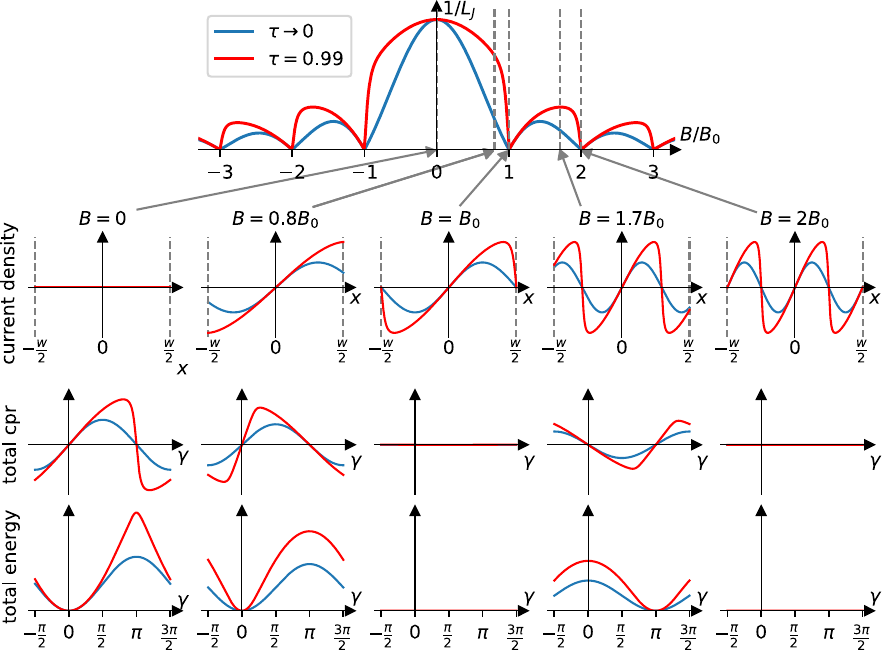}
	\caption{ Illustration of the interference effects in a wide junction. The upper panel shows the resulting interference pattern of the inverse Josephson inductance. The different rows show the current density, total CPR and total potential corresponding to different values of the out-of-plane field $B$ in each column, as indicated with the arrows and dashed gray lines in the upper panel.   }
	\label{fig:fraunhofer2}
\end{figure} 
It is assumed that the local CPR is given by $I(\phi)$. The varying phase leads to the $x-$dependent current density $j(x,\gamma) = \tfrac{1}{w}I(\phi(x,\gamma))$ along the width of the junction. The total CPR is obtained by integrating along the junction width (Eq. \ref{eq:integration_total_current}).
This integration corresponds to summing up CPRs which are shifted relative to each other. In the case of a sinusoidal CPR, the resulting total CPR $I_\Sigma(\gamma)$ is still sinusoidal. However, the shape of the total CPR can differ from $I(\gamma)$ if the CPR is non-sinusoidal. This can be seen in Fig. \ref{fig:fraunhofer2}. The critical current and Josephson inductance can be determined as usual by finding the maximum of the CPR and taking the derivative with respect to $\gamma$, respectively:
\begin{equation}
	L_J=\frac{\hbar}{2e} \left(\pdv{I_\Sigma}{\gamma}\Big|_{\gamma_0} \right)^{-1},
\end{equation}
with $\gamma_0$ the equilibrium phase (i.e. the phase where the potential is minimal). Its value depends on the choice of the integration boundaries in equation (\ref{eq:integration_total_current}). To simplify this expression, the following identity is needed:
\begin{align}
	\frac{\mathrm d}{\mathrm dx}f(ax+by_0)\Big|_{x=x_0} = f'(ax_0+by_0)\cdot a =\\ =\frac{\mathrm d}{\mathrm dy} f(ax_0+by)\Big|_{y=y_0}\cdot \frac{a}{b},
\end{align}
where $f'(\xi_0) := \pdv{f(\xi)}{\xi}\Big|_{\xi=\xi_0}$ the derivative of $f$ evaluated at $\xi_0$. For readability we introduce $\alpha = \frac{2\pi B_\perp}{B_0}$, such that $\phi(x,\gamma) = \gamma+\alpha x/w$. It follows
\begin{align}
	\frac{\mathrm dI_\Sigma}{\mathrm d\gamma}\Big|_{\gamma_0} &= \frac{\mathrm d}{\mathrm d\gamma} \frac{1}{w}\int_{-\tfrac{w}{2}}^{\tfrac{w}{2}}\!\mathrm d{x}\, I(\gamma+\alpha \tfrac{x}{w}) \Big|_{\gamma_0} =\\ &= \frac{1}{w}\int_{-\tfrac{w}{2}}^{\tfrac{w}{2}}\!\mathrm d x\,\frac{\mathrm d}{\mathrm d\gamma}I(\gamma+\alpha \tfrac{x}{w}) \Big|_{\gamma_0} = \\
	&= \frac{1}{\alpha}\int_{-\tfrac{w}{2}}^{\tfrac{w}{2}}\!\mathrm d{x}\, \frac{\mathrm d}{\mathrm dx}I(\gamma_0+\alpha \tfrac{x}{w}) =\\&= \frac{1}{\alpha}\left(I\left( \gamma_0+\tfrac{\alpha}{2} \right)-I\left( \gamma_0-\tfrac{\alpha}{2} \right)\right).
\end{align}
If we assume that the CPR is anti-symmetric around $\gamma_0$, as is the case for the CPR given by equation \ref{eq:beenakker}, then
\begin{align}
	\pdv{I_\Sigma}{\gamma}\Big|_{\gamma_0} &= \frac{2}{\alpha}I\left(\gamma_0 + \tfrac{\alpha}{2}\right) = \frac{I\left(\gamma_0+\frac{\pi B_\perp}{B_0}\right)}{\frac{\pi B_\perp}{B_0}}\\
	\Rightarrow\ L_J &= \frac{\hbar}{2e}\frac{\frac{\pi B_\perp}{B_0}}{I\left(\gamma_0+\frac{\pi B_\perp}{B_0}\right)} \label{eq:interference_Lj}.
\end{align}

For the CPR given by equation \ref{eq:beenakker}, there are two solutions which fulfill the condition $I_\Sigma(\gamma_0) = 0$ up to periodicity: $\gamma_0 = 0$ and $\gamma_0 = \pi$. This is because the CPR is an odd function around these points. Depending on the value of the magnetic field, one of the solutions corresponds to a potential minimum, the other to a potential maximum. We choose the value of the potential minimum. Within the central lobe of the Fraunhofer pattern, this is the case for $\gamma_0 = 0$, within the first side-lobe of the pattern, this is the case for $\gamma_0=\pi$, and so on. Hence the shape of the pattern within the central lobe is given by 
\begin{equation}\label{eq:interference_firstLobe}
	L_J = \frac{\hbar}{2e}\frac{\frac{\pi B_\perp}{B_0}}{I\left(\frac{\pi B_\perp}{B_0}\right)}.
\end{equation}
Fig \ref{fig3}a-c show the interference pattern in critical current and Josephson inductance for different values of the transparency. \\
For the critical current interference pattern, we could not find an analytic expression. The integral in Eq. (\ref{eq:integration_total_current}) is therefore computed numerically. We find that for a Beenakker-Furusaki CPR, the interference pattern shows no visible deviation from the sinc-shape, independent of the transparency $\tau$.

\section{Fitting the interference pattern}\label{ap:fittingTau}
\setcounter{figure}{0}

\begin{figure}[h!]
	\centering
	\includegraphics[width=1\columnwidth]{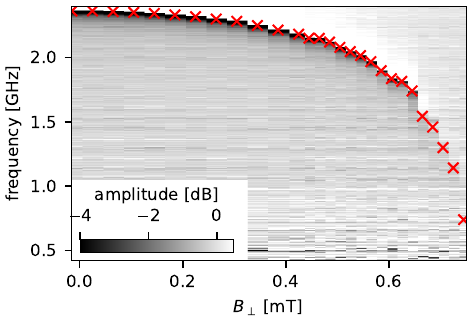}
	\caption{Colorplot of the first lobe of the interference pattern of device A. The background is subtracted using a reference measurement at higher temperature and magnetic field. The red crosses indicate the resonance frequencies extracted by a fit of the transmission traces.}
	\label{fig:appendix_fraunhoferFits}
\end{figure}

\begin{figure}[h!]
	\centering
	\includegraphics[width=1\columnwidth]{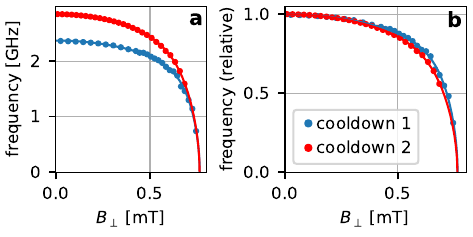}
	\caption{Comparison of the interference pattern of the same device A, measured in two different cooldowns. The sample was exposed to ambient conditions for approximately 120 days in between the two cooldowns. The dots represent the measured resonance frequencies, and the lines show the fit to Eq. (\ref{eq:LjInterferencePattern}) and Eq. (\ref{eq:beenakker}). For cooldown 1, a transparency of $\tau\approx 0.94$ was extracted, while cooldown 2 yielded a slightly lower value of $\tau\approx 0.896$. } 
	\label{fig:appendix_2ndCD}
\end{figure}

In the first step, the $\mathrm S_{21}$ traces of the central lobe are fitted using the formula given in \cite{deng-2013} in order to extract the resonance frequencies (red crosses in Fig. \ref{fig:appendix_fraunhoferFits}). For this, the background of the transmission measurement was subtracted using a reference measurement at higher temperature and magnetic field. 
The extracted interference pattern is then fitted using Eq.~(\ref{eq:LjInterferencePattern}), with the Beenakker-Furusaki CPR given in Eq.~(\ref{eq:beenakker}). The fit parameters are the junction transparency $\tau$ and the magnetic field periodicity $B_0$. The error in the final value of the transparency given in the main-text is mainly due to the uncertainty of the offset field. The exact value of the offset field is not known as we could not perform a compensation measurement on the same chip. We are assuming an uncertainty of $\pm 15\,\mathrm{\mu T}$. The resulting uncertainty of the transparency was determined by fitting the pattern for extreme values of the offset field. \\
As demonstrated in the main-text, the CPR can be directly extracted from the measured interference pattern by inverting Eq. (\ref{eq:LjInterferencePattern}):
\begin{equation}
    I(\phi) = \frac{\hbar}{2e}\frac{\phi}{L_J(B_\perp=B_0\phi/\pi)}.
\end{equation}
The magnetic field $B_\perp$ is converted to a phase difference $\phi$ by a factor $B_0/\pi$. Therefore, the fitted value of $B_0$ is required to establish the phase scale of the extracted CPR from the measured data. The resulting CPR is shown in Fig. \ref{fig3}e.\\
Device A was cooled down twice. In the second cooldown, the transparency was slightly lower, as illustrated in Fig. \ref{fig:appendix_2ndCD}.

\begin{acknowledgments}
We thank Alfredo Levy Yeyati and Francisco Matute-Cañadas for useful discussions and comments on the manuscript. We acknowledge the support of Julius K\"uhne and Ian D. Sharp from Walter Schottky Institute for the ALD growth. The work in Regensburg was funded by the Deutsche Forschungsgemeinschaft (DFG, German Research Foundation) within Project-ID 314695032-SFB 1277 and INST-89/561-1 FUGG. N.P. and C.S. acknowledge funding by EU’s HORIZON-RIA Programme under Grant No. 101135240 (JOGATE). L.T. acknowledges the Georg Forster Fellowship from the Humboldt Foundation. L.T. and A.K. thank the support of the COST Action CA21144 Program. C.S. acknowledges the DFG as part of the German Excellence Strategy – EXC3112/1 – 533767171 (Center for Chiral Electronics). D.B. and C.S. acknowledge the Munich Quantum Valley program, which is supported by the Bavarian state government with funds from the Hightech Agenda Bavaria. G.B. acknowledges funding by EU's Horizon-Pathfinder Programme under Grant No. 101115315 (QuKiT). D.C. acknowledges funding by Next Generation EU M4C2, PNRR project NFFA-DI, CUP B53C22004310006, IR0000015. This project has received funding from the European Union’s  Horizon 2020 research and innovation programme under grant agreement No 101007417 within the framework of the NFFA-Europe Pilot Transnational Access Activity, proposal ID897, having benefited from the access provided by CNR IOM.
\end{acknowledgments}

\bibliography{bib/bib-alex}
\end{document}